\documentclass{revtex4-1}
\usepackage{graphicx}
\usepackage{dcolumn}
\usepackage{bm}
\usepackage[T1]{fontenc}
\usepackage[english]{babel}
\usepackage{amsmath}
\usepackage{amssymb}
\usepackage{tabularx}
\usepackage{multirow}
\usepackage[all]{xy}
\usepackage{appendix}
\usepackage[dvips]{color}

\begin{document}

\title{
Transition from plasma- to Kerr-driven laser filamentation}

\author{P. B\'ejot$^{1,2}$}\email{pierre.bejot@unige.ch}
\author{E. Hertz$^{1}$}
\author{J. Kasparian$^{2}$}\email{jerome.kasparian@unige.ch}
\author{B. Lavorel$^{1}$}
\author{J.-P. Wolf$^{2}$}
\author{O. Faucher$^{1}$}

\affiliation{(1) Laboratoire Interdisciplinaire Carnot de Bourgogne (ICB), UMR 5209 CNRS-Universit\'e de Bourgogne, BP 47870, F-21078 Dijon Cedex, France}
\affiliation{(2) GAP-Biophotonics, Universit\'e de Gen\`eve, 20 rue de l'\'{E}cole de M\'edecine, 1211 Geneva 4, Switzerland}

\begin{abstract}
While filaments are generally interpreted as a dynamic balance between Kerr focusing and plasma defocusing, the role of the higher-order Kerr effect (HOKE) is actively debated as a potentially dominant defocusing contribution to filament stabilization. In a pump-probe experiment supported by numerical simulations, we demonstrate the transition between two distinct filamentation regimes at 800\,nm.  For long pulses (1.2\,ps), the plasma substantially contributes to filamentation, while this contribution vanishes for short pulses (70\,fs). These results confirm the occurrence, in adequate conditions, of filamentation driven by the HOKE rather than by plasma.
\end{abstract}

\maketitle

Filamentation \cite{BraunKLDSM1995,CouaironM07,Chin,Kasparian} is a self-guided propagation regime typical of high-power lasers, offering spectacular potential applications \cite{Kasparian_Opex} like rainmaking \cite{Rain_Nature} and lightning control \cite{Lightning_Opex}.
We recently challenged its long-established mechanism by measuring the higher order Kerr effect (HOKE) in gases, implying that the non-linear refractive index must be written as $\Delta n_{\rm{Kerr}}=\sum{n_{2j}I^j}$, where the non-linear indices $n_{2j}$ are related to the $(2j+1)^{th}$ electric susceptibility $\chi^{(2j+1)}$. The inversion of  $\Delta n_{\rm{Kerr}}$, due to negative $n_4$ and $n_8$ terms in air and argon, leads to a defocusing Kerr effect at an intensity close to those present within filaments \cite{Loriot09,arXivLoriot}. As a consequence, the HOKE can ensure self-defocusing in filaments and balance Kerr self-focusing \cite{Bejot}, in place of the plasma, especially for short pulses  \cite{arXivLoriot}. 
 This result raised an active controversy \cite{Kolesik1,Kolesik2,ChenVAM2010,SteinBD2010,HarmonicFaucher,KosarDPWHYRMC2011} in the lack of direct experimental confirmation. 

\emph{Quantitative} differences between the predictions of filamentation models including or disregarding the HOKE are not sufficient to distinguish between them.
The intensity within filaments ($\sim$50 TW/cm$^2$ \cite{BeckeAVOBC2001}) is compatible with regularization by either the plasma \cite{Kasparian2} or 
{the HOKE} \cite{Bejot}, which balance the Kerr self-focusing in the same intensity range. Furthermore the electron density is  difficult to measure and highly dependent on initial conditions, resulting in a wide spread of the reported values from 10$^{12}$ to 10$^{17}$ cm$^{-3}$ \cite{Kasparian2},
{although} the latest measurements range from $\sim$10$^{15}$ cm$^{-3}$ \cite{ThebergeLSBC06} to a few 10$^{16}$ cm$^{-3}$ \cite{ChenVAM2010}.


In the present Letter, we therefore focus on experimental conditions where  \emph{qualitatively different behaviors} of plasma- and HOKE-driven filamentation allows to unambiguously distinguish between them. This discrimination proceeds from the intrinsically different temporal dynamics of these non-linear defocusing contributions. While the Kerr effect is instantaneous at the timescale of the pulse envelope, the free electrons accumulated throughout the pulse duration {survive} for tens of picoseconds after the laser pulse has passed \cite{TzortPFM2000,Loriota}. In an atomic gas like argon, where no spatio-temporal modification of the refractive index due to molecular alignment occurs, 
two pulses separated by a few picoseconds can therefore only be coupled if the free electron density left by the first one is sufficient to affect the second one. This  allows to distinguish between two scenarios. If the plasma is the dominant mechanism for the self-guiding of the pump pulse, then the electron density produced in its wake is necessarily affect the probe filament. On the contrary, if the HOKE terms are predominant, then the probe filament is insensitive to the presence of the pump. 

Based on these different temporal dynamics, we unambiguously observe experimentally the all Kerr-driven filamentation of ultrashort laser pulses, as well as the transition to a plasma-driven filamentation regime in the case of long pulses.
This new perspective on the filamentation physics critically revises the discussion of the optimal laser parameters for applications ranging from laser-controlled atmospheric experiments to harmonics generation \cite{harmoniques,Lhuillier24_1991}.

\begin{figure} [b!]
   \begin{center}
      \includegraphics[keepaspectratio,width=8.6cm]{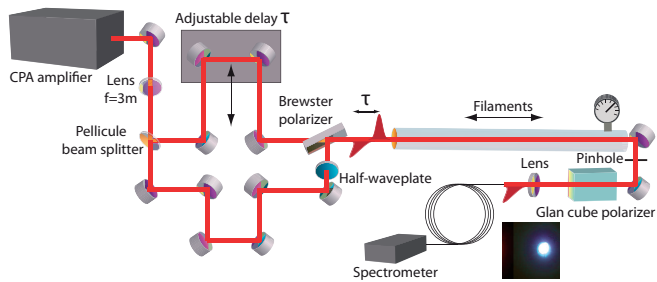}
   \end{center}
   \caption{Experimental setup. Two orthogonally polarized identical pulses with an adjustable delay are focused in an argon-filled cell. At the output of the cell, the probe spectrum is selected by a Glan cube polarizer and analyzed with a spectrometer.}
\label{Fig:1}
\end{figure}


In our experimental setup (Fig.~\ref{Fig:1}), two orthogonally polarized co-propagating laser pulses centered at 800\,nm, are loosely focused by a $f$=3\,m lens in a 2\,m-long gas cell filled with argon. We investigate both short pulses (70\,fs, 600\,$\mu$J, 3\,bar) and long pulses (1.2\,ps, 6\,mJ, 5\,bar), keeping the peak power of each pulse equal to 2.5 critical powers. At this power level, each pulse produces a single filament when propagating alone. The probe pulse is temporally delayed with regard to the pump by $\tau$ ($\tau$=1\,ps for short pulses and $\tau$=2\,ps for long pulses).

The influence of the plasma left by the pump on the probe {filament} is characterized by {observing changes in its spectrum. The filament output spectrum is selected in the far-field ($z\sim$2 m) by a pinhole excluding the conical emission and most of the photon bath}, and {analyzed} with a spectrometer (Ocean Optics HR4000, providing 0.5 nm resolution {and 14 bits of dynamic range}) after {separating} the two pulses at the cell exit using a Glan-Thomson polarizer. {To improve the signal-to-noise over the whole considered spectral range, each spectrum is reconstructed by assembling data from 3 spectral ranges. For each range, the integration time was adjusted between 2 and 2000 pulses to ensure the use of the full dynamics of the spectrometer in every spectral region. The resulting spectra were then averaged over 20 realizations.} 
The change in the probe spectra induced by the pump pulse is characterized by calculating the contrast $C(\lambda)=(S_1(\lambda)-S_0(\lambda))/(S_1(\lambda)+S_0(\lambda))$ between the spectral densities with ($S_1$) and without ($S_0$) the pump pulse at each wavelength $\lambda$.

Since the spatial overlap all along the propagation is crucial for the relevance of the measurements, it is optimized by maximizing the interference pattern produced by the unfocused pulses both before and after the cell. It is also confirmed by the occurrence of multiple filamentation at zero delay, which is set by optimizing frequency doubling in a BBO crystal placed before the cell. 
Moreover, we checked that the alignment is conserved when translating the probe pulse from $\tau$=0\,ps to $\tau$=1\,ps (resp. $\tau$=2\,ps) by inserting a 200\,$\mu$m (resp. 400\,$\mu$m) thick glass window in the path of the short (resp. long) pump pulse, delaying it by 1\,ps (resp. 2\,ps) and checking that the multi-filamentation is restored. {
Let us note that the action of the pump pulse on the probe filaments may induce a longitudinal or transverse displacement. Such coupling, however, would mean that the plasma strongly affect the filamentation dynamics close to the non-linear focus, where a substantial part of the white-light continuum is generated. It would therefore result in significant changes in the output spectra.}
Finally, no broadening is recorded in vacuum, confirming that neither input nor output windows of the cell have significant contributions on the spectral broadening.

\begin{figure} [!]
   \begin{center}
      \includegraphics[keepaspectratio,width=8.6cm]{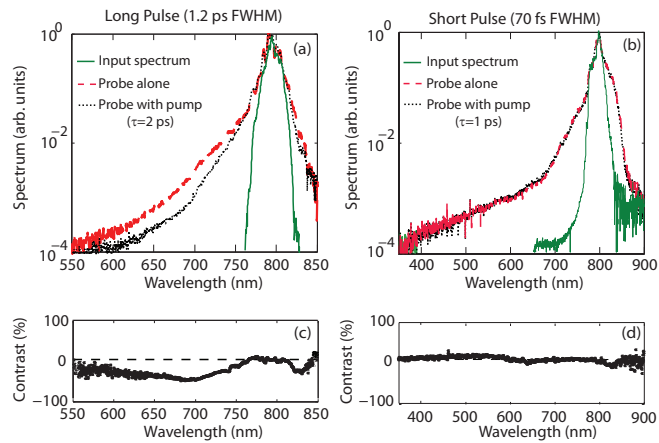}
   \end{center}
   \caption{(a,b) Experimental spectra of the filaments generated by a long (1.2 ps, a) and short (70 fs, b) pulse conveying 2.5 critical powers, both with and without a pump pulse $\tau$=2~ps and $\tau$=1~ps ahead of it, respectively. This delay ensures that the pump pulse can only influence the probe pulse through the defocusing free electrons {left} behind by the pump.
 The input spectra are also given for reference. (c-d) Contrast between the spectra with and without pump pulse. The probe pulse is affected by the plasma left by the pump pulse only in the long-pulse regime.}
   \label{Fig:2}
\end{figure}

As shown in Fig.\,\ref{Fig:2}a, the spectrum of the filament generated by a 1.2 ps long pulse narrows when it is preceded by a pump filament $\tau$=2\,ps ahead of it. 
This coupling demonstrates that the probe pulse propagates through a pre-ionized medium with a free electron density providing a significant negative contribution to the refractive index, i.e. a non-negligible defocusing term: As expected from the Kerr-plasma balance model, the plasma density generated in the filaments substantially contributes to the self-guiding. In this framework, the narrower spectrum of the probe filament is easily explained by the supplementary plasma density left by the pump pulse, which decreases the clamping intensity ensuring the balance in the probe pulse between Kerr self-focusing and defocusing by the plasma, and consequently reduces the 
{efficiency} of its spectral broadening.

Conversely, in the case of short pulses (70 fs, Fig.\,\ref{Fig:2}b), the spectrum of the probe is
{nearly}  insensitive to the plasma left by the {pump.}
The decoupling between the two pulses separated by a delay ($\tau=$1\,ps) too short to allow electron recombination unambiguously shows that the plasma density left in the wake of the pump pulse is too weak to significantly affect the filamentation process in the considered conditions. Plasma therefore plays no role in filamentation of 70 fs pulses, which is instead driven by the HOKE as predicted numerically \cite{Bejot}.
Note that, if the two pulses overlap temporally ($\tau$=0), their coupling is restored due to cross-phase modulation
, confirming that the two pulses indeed overlap {both longitudinally and transversely}. 
These results therefore provide the experimental demonstration of the transition from plasma-driven filamentation in the case of long pulses to HOKE-driven filamentation for shorter pulses, as expected from theoretical considerations \cite{arXivLoriot}.


In order to support this qualitative discussion and provide a closer look on the physical process at play, we numerically investigate the propagation of the pulses in the present experimental conditions. In a first step, we simulate the propagation of the pump pulse. 
The plasma density left behind this pulse is then used as an initial condition for calculating the probe pulse propagation. 

The numerical model considers  linearly polarized incident electric fields at a wavelength $\lambda_0$=$800$ nm with cylindrical symmetry around the propagation axis $z$. According to the unidirectional propagation pulse equation \cite{Moloney_UPPE}, the scalar envelope $\varepsilon(r,t,z)$ (defined such that $|\varepsilon(r,z,t)|^2=I(r,z,t)$, $I$ being the intensity) evolves in the frame traveling at the pulse velocity according to:
\begin{eqnarray}
\begin{aligned}
\partial_z\widetilde{\varepsilon}=&i(\sqrt{k^2(\omega)-k_\bot^2}-k'\omega)\widetilde{\varepsilon}&\\
&+\frac{1}{\sqrt{k^2(\omega)-k_\bot^2}}\left(\frac{i\omega^2}{c^2}\widetilde{P}_{\textrm{NL}}-\frac{\omega}{2\epsilon_0c^2}\widetilde{J}\right)-\widetilde{\alpha},&
\label{Eq.final_enveloppe}
\end{aligned}
\end{eqnarray}
where $c$ is the velocity of light in vacuum, $\omega$ is the angular frequency, $k(\omega)$=n($\omega$)$\omega$/c, $k'$ its derivative at $\omega_0$=2$\pi$c/$\lambda_0$, $n(\omega)$ is the linear refractive index at the frequency $\omega$, $k_{\bot}$ is the spatial angular frequency. $P_{\textrm{NL}}$ is the nonlinear polarization, $J$ is the free-charge induced current and $\alpha$ is the nonlinear losses induced by photo-ionization. $\widetilde{f}$ denotes simultaneous temporal Fourier and spatial Hankel transforms of function $f$: $\widetilde{f}=\int_0^{\infty}{\int_{-\infty}^{\infty}r J_0(k_\bot r) f(r,t)e^{i\omega t}dtdr}$, where $J_0$ is the zeroth order Bessel function and $f\equiv\varepsilon, {P}_{\textrm{NL}}, J$, or $\alpha $. $P_{\textrm{NL}}$ is evaluated as $P_{\textrm{NL}}=\sum{n_{2j}|\varepsilon|^{2j}}\varepsilon$, where $n_{2j}$ are the $j^{\textrm{th}}$-order nonlinear refractive indices as measured in \cite{Loriot09}. The current is evaluated as $\widetilde{J}=\frac{e^2}{m_e}(\nu_e+i\omega)\widetilde{\rho\varepsilon}/(\nu_e^2+\omega^2)$, where $e$ and $m_e$ are the electron charge and mass respectively, $\epsilon_0$ is the vacuum permittivity, $\nu_e$ is the effective electronic collisional frequency, and $\rho$ is the {electron} density. Finally, $\alpha=W(|\varepsilon|^2)U_i\left(\rho_{\textrm{at}}-\rho\right)/(2|\varepsilon|^2)$, $\rho_{\textrm{at}}$ is the neutral atoms density, $W(|\varepsilon|^2)$ is the photoionization probability modeled by the PPT (Perelomov, Popov, Terent'ev) formulation, with ionization potential $U_i$.

The propagation dynamics of the electric field is coupled with the electron density $\rho$, calculated as \cite{CouaironM07}
\begin{equation}
\partial_t\rho=W(|\varepsilon|^2)\left(\rho_{at}-\rho\right)+\frac{\sigma}{U_i}\rho|\varepsilon|^2-\beta\rho^2,
\label{Eq.electron_dens}
\end{equation}
where $\beta$ is the electron recombination rate and $\sigma$ is the inverse Bremsstrahlung cross-section of argon{, also accounting for avalanche ionization. The output spectrum is integrated over 2 mrad around the beam center to match the experimental conditions}.

\begin{figure} [!]
   \begin{center}
      \includegraphics[keepaspectratio,width=8.6cm]{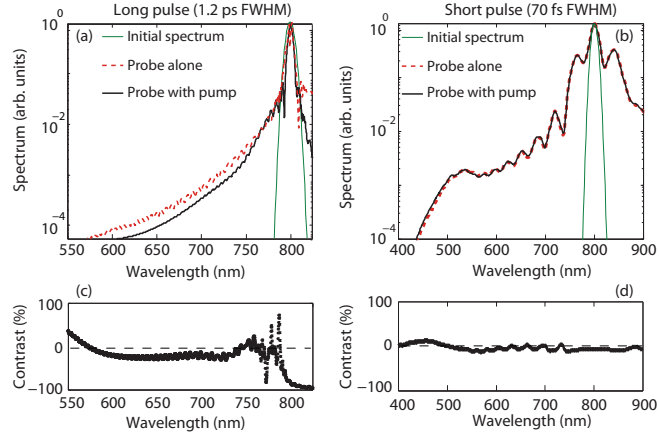}
   \end{center}
   \caption{(a,b) Spectra of the filaments generated by the long (a) and short (b) pulse, simulated by the full numerical model considering the HOKE. (c-d) Contrast between the spectra with and without pump pulse. The probe pulse is affected by the plasma left by the pump pulse only in the long-pulse regime.}
\label{Fig:3}
\end{figure}

\begin{figure} [!]
   \begin{center}
      \includegraphics[keepaspectratio,width=8.6cm]{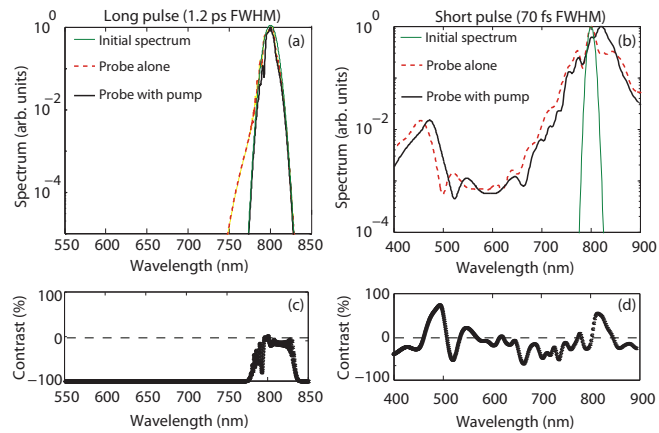}
   \end{center}
   \caption{(a,b) Spectra of the filaments generated by the long (a) and short (b) pulse, simulated by the truncated numerical model disregarding the HOKE. (c-d) Contrast between the spectra with and without pump pulse. The probe pulse is affected by the plasma left by the pump pulse whatever the pulse duration.}
\label{Fig:4}
\end{figure}

The full model including the contribution of the HOKE together with the plasma response reproduces remarkably well the experimentally observed behavior. In particular, as displayed in Fig.~\ref{Fig:3}, the spectral broadening of the long probe pulse is reduced when the pump pulse precedes it, in a ratio comparable with the experimental data, while the filament generated by the 70 fs probe pulse is unaffected by the pump pulse.
In contrast, disregarding the HOKE (Fig.~\ref{Fig:4}) would lead us to expect that the filament output spectrum generated by probe pulses of any duration should be affected by the presence of the pump pulse, in contradiction with the experimental results. Furthermore, this truncated model inadequately predicts the wings of the spectral broadening, even for a single pulse. Comparing both Figs.~\ref{Fig:3} and \ref{Fig:4} with Fig.~\ref{Fig:2} clearly illustrates the need to consider the HOKE in numerical simulations of laser filamentation, even in the plasma-driven, long-pulse filamentation regime. 

This need is illustrated by comparing the electron densities predicted by both models. While the truncated model yields 10$^{17}$ cm$^{-3}$ for both 1.2 ps and 70 fs pulses, the full model 
yields 4$\times$10$^{16}$ cm$^{-3}$ and 10$^{15}$ cm$^{-3}$, respectively. This strong dependence of the plasma contribution on the pulse duration explains both the contrasted behaviors observed in our experiment between the plasma- and HOKE-driven filamentation regimes, but also the wide spread of the experimentally measured electron densities in filaments \cite{Kasparian2,ThebergeLSBC06,ChenVAM2010}. Furthermore, we can estimate the relative contributions of HOKE and plasma to defocusing by considering the ratio $\xi=\Delta n_{\rm{HOKE}}/\Delta n_{\rm{plasma}}$ of the HOKE- to plasma-induced refractive index change. 
For short pulses (70 fs), this ratio keeps well above {1} all along the filament length  ($\xi \ge 39$), illustrating the negligible contribution of plasma to the filamentation process. In contrast, for 1.2 ps this ratio keeps close to 1 ($\xi \gtrsim 0.72$, except for a spike with $\xi=0.24$ at the very filament onset), confirming that, while plasma provides the major defocusing contribution, the HOKE are far from negligible even in these conditions.



As a conclusion, a pump-probe experiment 
allowed us to unambiguously observe experimentally the theoretically predicted HOKE-driven filamentation {for ultrashort pulses} \cite{Bejot}, as well as the transition from this regime
 to the long-known plasma-driven filamentation regime for long pulses \cite{arXivLoriot}. This transition is similar to that observed in the context of high-harmonic generation (HHG), where the use of too long pulses results in gas ionization instead of HHG \cite{Lhuillier24_1991}. 
Furthermore, comparing our results with numerical simulations show that, even in the plasma-driven filamentation regime {of the present work}, the contribution of the HOKE to the propagation dynamics cannot be neglected. 
This finding provides a better understanding of filamentation, and therefore allows to improve its modelling. It further confirms the relevance of the measured HOKE \cite{Loriot09,arXivLoriot}, with implications ranging from spectral broadening in optical fibers \cite{PopmiCAMK2010} to the generation of few-cycle pulses \cite{HauriKHCMBK2004}, 
atmospheric applications \cite{Kasparian_Opex,Rain_Nature,Lightning_Opex}, or fermionic light \cite{NovoaMT2010}. 

%
%
%
%
\textbf{Acknowlegments.}
This work was supported by the ANR COMOC ANR-07-BLAN-0152-01, FASTQUAST ITN Program of the seventh FP and the Swiss NSF (contract number 200021-125315). We also acknowledge L.\,Bonacina, A.\,Rondi, J.\,Extermann, and M. Moret for their experimental support.

\bibliographystyle{unsrt}

\end{document}